Литвинова С. Г.
*Інститут інформаційних технологій і засобів навчання НАПН України*
*s.h.lytvynova@gmail.com*
*ORCID (http://orcid.org/0000-0002-5450-6635)*
*DOI 10.31110/2413-1571-2019-019-1-017*

## МОДЕЛЬ ВИКОРИСТАННЯ СИСТЕМИ КОМП'ЮТЕРНОГО МОДЕЛЮВАННЯ ДЛЯ ФОРМУВАННЯ КОМПЕТЕНТНОСТЕЙ УЧНІВ З ПРИРОДНИЧО-МАТЕМАТИЧНИХ ПРЕДМЕТІВ

***Анотація.***

***Формулювання проблеми.*** *Відсутність методичного забезпечення, низький рівень обізнаності вчителів щодо існуючих ефективних технологій навчання таких, як комп'ютерне моделювання не дає змоги учням формувати індивідуальну траєкторію розвитку і формувати компетентності з природничо-математичної освіти. Це щорічно підтверджується аналітичними звітами Українського центру оцінювання якості освіти, зокрема оприлюднюються результати зовнішнього незалежного оцінювання учнів з природничо-математичних предметів.*

***Матеріали і методи.*** *У процесі дослідження використовувались методи аналізу педагогічної, методичної літератури і дисертаційних досліджень; здійснювалося узагальнення результатів вітчизняного і зарубіжного досвіду; теоретичне моделювання процесу використання системи комп'ютерного моделювання для формування компетентностей учнів; системний аналізу для визначення структурних елементів моделі системи комп'ютерного моделювання.*

***Результати.*** *Обґрунтовано і розроблено модель використання системи комп'ютерного моделювання для формування компетентностей учнів з природничо-математичних предметів. До особливостей цієї моделі можна віднести її використання на засадах дитиноцентриського, системного, діяльнісного, диференційованого підходів з метою формування компетентностей учнів з природничо-математичних предметів. Важливим результатом дослідження є обґрунтування: критеріїв добору СКМод (варіативність, доступність, інструментальність, інтерактивність, інтуїтивність, комплексність, наочність даних, науковість моделей, незалежність, реалістичність); форм і методів застосування СКМод в освітньому процесі, виокремлення аспекту застосування пізнавальних завдань (дослідницьких, творчий, прикладних) для формування компетентностей учнів; обґрунтування таких форм оцінювання як рефлексія, формуюче оцінювання, підсумкове оцінювання.*

***Висновки.*** *Системи комп'ютерного моделювання стають альтернативою звичайним малюнкам у паперових підручниках і плакатам. Імітаційні процеси, інтерактивні вправи різних типів, анімаційний супровід, формуюче оцінювання дають можливість кожному учню формувати компетентності і пізнавати світ з захопленням. Використання СКМод урізноманітнює процес здобуття освіти, дозволяє перейти від пасивних до активних методів навчання, активізує навчально-пізнавальну діяльність учнів, дозволяє створити індивідуальну траєкторію розвитку для кожного учня у процесі вивчення природничо-математичних предметів.*

*Подальшого обґрунтування потребує розроблення методичних рекомендацій щодо застосування СКМод в освітньому процесі, що слугуватимуть науково-методичним забезпеченням формування компетентностей учнів з природничо-математичних предметів.*

***Ключові слова:*** *система комп'ютерного моделювання; СКМод; моделювання; заклади загальної середньої освіти; природничо-математична освіта; компетентність; пізнавальні завдання.*

### 1. ВСТУП

**Постановка проблеми.** У Білій книзі національної освіти України зазначено, що, уже з'явився новий напрям діяльності – розроблення ІКТ навчання і програмно-методичних навчальних комплексів, що базуються на широкому застосуванні інтерактивних методів навчання, мультимедійних засобів і віртуальних педагогічних технологій, які дають змогу суттєво підвищити рівень методичного забезпечення освітнього процесу, відкривають нові можливості для підвищення якості освіти. Тому інформатизація і комп'ютеризація є одним із найважливіших і водночас найскладніших сучасних завдань галузі освіти (Кремень, 2009).

Зміни в освіті обумовлені загальним рівнем розвитку суспільства; формуванням нової генерації людей, які живуть в середовищі, насиченому цифровими засобами; появою нових технологій, що потребують постійного доступу до мережі Інтернет. Технологічні зміни, що відбулися у формуванні освітнього середовища закладів загальної середньої освіти, дали можливість учителям надавати учням приклади і зразки пошукової, творчої діяльності, пізнавальні завдання з використання ІКТ, зокрема систем комп'ютерного моделювання (СКМод) (Литвинова, 2018; Пінчук&Литвинова&Буров, 2017).

Відсутність методичного забезпечення, низький рівень обізнаності вчителів щодо існуючих ефективних технологій навчання таких, як комп'ютерне моделювання не дає змоги учням формувати індивідуальну траєкторію розвитку і формувати компетентності з природничо-математичної освіти (Литвинова, 2018). Це щорічно підтверджується аналітичними звітами Українського центру оцінювання якості освіти, зокрема оприлюднюються результати зовнішнього незалежного оцінювання учнів з природничо-математичних предметів.

**Аналіз актуальних досліджень.** Питання ефективності використання комп'ютерного моделювання в освітньому процесі досліджувалися протягом останніх років в роботах вітчизняних вчених. Одним із основних питань для впровадження комп'ютерного моделювання в освітній процес було обґрунтування проблеми формування системи відкритої освіти, використання інноваційних середовищ навчання – комп'ютерно-орієнтованих і мобільних, електронних освітніх ігрових ресурсів та імітаційних моделей (Биков, 2009).

У роботі «Застосування ігрових симуляторів у формуванні професійних компетентностей майбутніх інженерів-програмістів» вченим проаналізовано низку сучасних ігрових симуляторів. Виділено критерії та показники добору ігрових симуляторів: дидактичний (відповідність темам та компетентностям; реалістичність; взаємодія з іншими ролями; можливість аналізу результатів та помилок; адаптивність рівня складності; підтримка різних сценаріїв); функціональний (зручність інтерфейсу; захоплюючий ігровий процес; безкоштовність; мультиплеєр; гра зі штучним інтелектом; багатомовність); технологічний (кросплатформеність; простота налаштування; сумісність з мобільними пристроями) (Концедайло, 2018).

Д. С. Антонюк у роботі «Використання програмно-імітаційних комплексів як засобів формування економічних компетентностей студентів технічних спеціальностей» обґрунтував та розробив авторську модель використання програмно-імітаційних комплексів, представив загальну структуру методики їх використання; описав різні форми та методи використання програмно-імітаційних комплексів в освітньому процесі (Антонюк, 2018).

У роботі «Методика використання комп'ютерного 3D проектування у навчанні майбутніх фахівців з дизайну» вчений обґрунтував етапи ознайомлення та оволодіння засобами ІКТ, що включають: узагальнено-оглядовий (передбачає теоретичне ознайомлення з основним функціоналом засобів ІКТ, можливостями та особливостями, рівнем застосування серед професіоналів), практично-оглядовий (передбачає теоретичне ознайомлення з практичним обмеженим оволодінням) та практичне застосування для виконання навчальних завдань (передбачає короткий теоретичний огляд та розширене практичне застосування в ході виконання циклу практичних та лабораторних завдань). Вчений також зазначає, що для ефективного навчання залишається необхідність застосування комплексного використання засобів ІКТ, серед яких електронна пошта, блог, «віртуальна дошка», соціальні мережі, офісні програми та додатки, графічні редактори, 3D-редактори (Борисенко, 2018). Результати наукового пошуку вченого щодо застосування комп'ютерного моделювання в освітньому процесі можна інтегрувати на різних рівнях освіти, зокрема в закладах загальної середньої освіти на уроках трудового навчання.

Питання комп'ютерного моделювання та застосування мережі Інтернет для дослідження природних явищ піднімалося співробітниками Інституту інформаційних технологій і засобів навчання НАПН України. Група дослідників обґрунтувала основні принципи і підходи використання Інтернет-технологій в шкільному експерименті під час вивчення курсу фізики. Вони довели, що для застосування комп'ютерного моделювання мають бути визначені конкретні навчальні цілі. За висновками вчених інноваційні підходи в навчанні спонукатимуть учнів до пошуку причинно-наслідкових зв'язків, допомагатимуть отримані знання і досвід, залучити до здобуття нових знань, співставлення цих знань з оточуючим світом (Жук&Соколюк&Дементієвська&Слободяник&Соколов, 2014).

В Університеті імені М.П. Драгоманова дослідники встановили, що засвоєння знань учнями відбувається більш ефективно в процесі діяльності. Такою діяльністю може бути розробка комп'ютерних моделей фізичних явищ. При цьому процес побудови комп'ютерної моделі можна організувати з поступовим її ускладненням «від простого до складного» (Матвійчук&Сергієнко&Подласов, 2008).

У Науково-дослідному інституті імені Вейцмана (Ізраїль) дослідники прийшли висновку, що використання комп'ютерного моделювання в освітньому процесі, сприяє розвитку інтелектуальних умінь учнів, глибокому розумінню природних процесів, формуванню дослідницьких навичок і вмінь, поглибленню знань з інформатичних, математичних та фізичних дисциплін, удосконаленню навичок роботи в цифровому середовищі (Хазіна, 2010).

Зазначимо, що питання використання комп'ютерних моделей в освітньому процесі розвивалося і досліджувалося вченими за такими напрямками: засвоєння базових предметів на засадах використання імітаційного моделювання (Павленко, 2011); використання імітаційного моделювання в освітньому процесі (Фадєєва, 2013); посилення міждисциплінарних зв'язків (Мястковська, 2014); підвищення інтересу учнів до навчання на засадах використання ігрового моделювання (Прокопенко, 2014); використання електронних освітніх ігрових ресурсів (Мельник, 2016; Рибалко, 2016); використання систем комп'ютерної математики GeoGebra з метою активізації дослідницької діяльності учнів (Гриб'юк, 2017); активізація пізнавальної діяльності учнів (Слободяник, 2018).

**Мета статті.** З огляду на це метою статті є обґрунтування моделі використання системи комп'ютерного моделювання в системі загальної середньої освіти для формування компетентностей учнів з природничо-математичних предметів.

**2. МЕТОДИ ДОСЛІДЖЕННЯ**

У процесі дослідження використовувались методи аналізу педагогічної і методичної літератури й дисертаційних досліджень; здійснювалося узагальнення результатів вітчизняного і зарубіжного досвіду; теоретичне моделювання використання системи комп'ютерного моделювання для формування компетентностей учнів; системний аналізу для визначення структурних елементів моделі системи комп'ютерного моделювання. Це дослідження виконувалося в рамках науково-дослідної роботи «Система комп'ютерного моделювання пізнавальних завдань для формування компетентностей учнів з природничо-математичних предметів» (НДР №0118U003160).

**3. РЕЗУЛЬТАТИ ДОСЛІДЖЕННЯ**

В умовах становлення і розвитку високотехнологічного інформаційного суспільства в Україні виникає необхідність підвищення якості та пріоритетності шкільної природничо-математичної освіти, включення природничо-математичних предметів до навчальних планів усіх рівнів освіти, поліпшення природничо-математичної підготовки учнів.

Метою природничо-математичної освіти є оволодіння суб'єктами навчального процесу знаннями з дисциплін, які характерні для цього напрямку, уміннями та навичками з метою подальшого їх використання у життєвій і професійній діяльності. Сьогодні одне з основних завдань природничо-математичної освіти є формування у молодого покоління цілісного природничо-наукового світогляду (Кохановська, 2015).

Розрізняють загальну та спеціальну природничо-математичну освіту. Загальна – забезпечує засвоєння сукупності знань з основ природничо-математичних дисциплін, які необхідні кожній людині незалежно від її професії. Спеціальна – систематизовані знання і практичні навички з природничих та математичних наук надаються в рамках здобуття майбутньої професії. Загальна природничо-математична освіта надається закладами загальної середньої освіти, спеціальна – забезпечується в середовищі закладів вищої освіти (Капров, 1965).

Нині викликає занепокоєння негативна тенденція, щодо зниження загального рівня підготовки випускників закладів загальної середньої освіти, зокрема щодо їхнього розуміння природних явищ і процесів, володіння знаннями необхідними для практичного застосування у процесі вирішення життєвих проблем. В учнів спостерігається відсутність цілісного уявлення про природу світу, що підтверджує низьку сформовану компетентність учнів з природничо-математичних предметів.

З метою поліпшення ситуації і підвищенням рівня компетентності учнів з природничо-математичних предметів пропонується модель використання системи комп'ютерного моделювання та пізнавальних завдань в системі загальної середньої освіти (рис. 1).

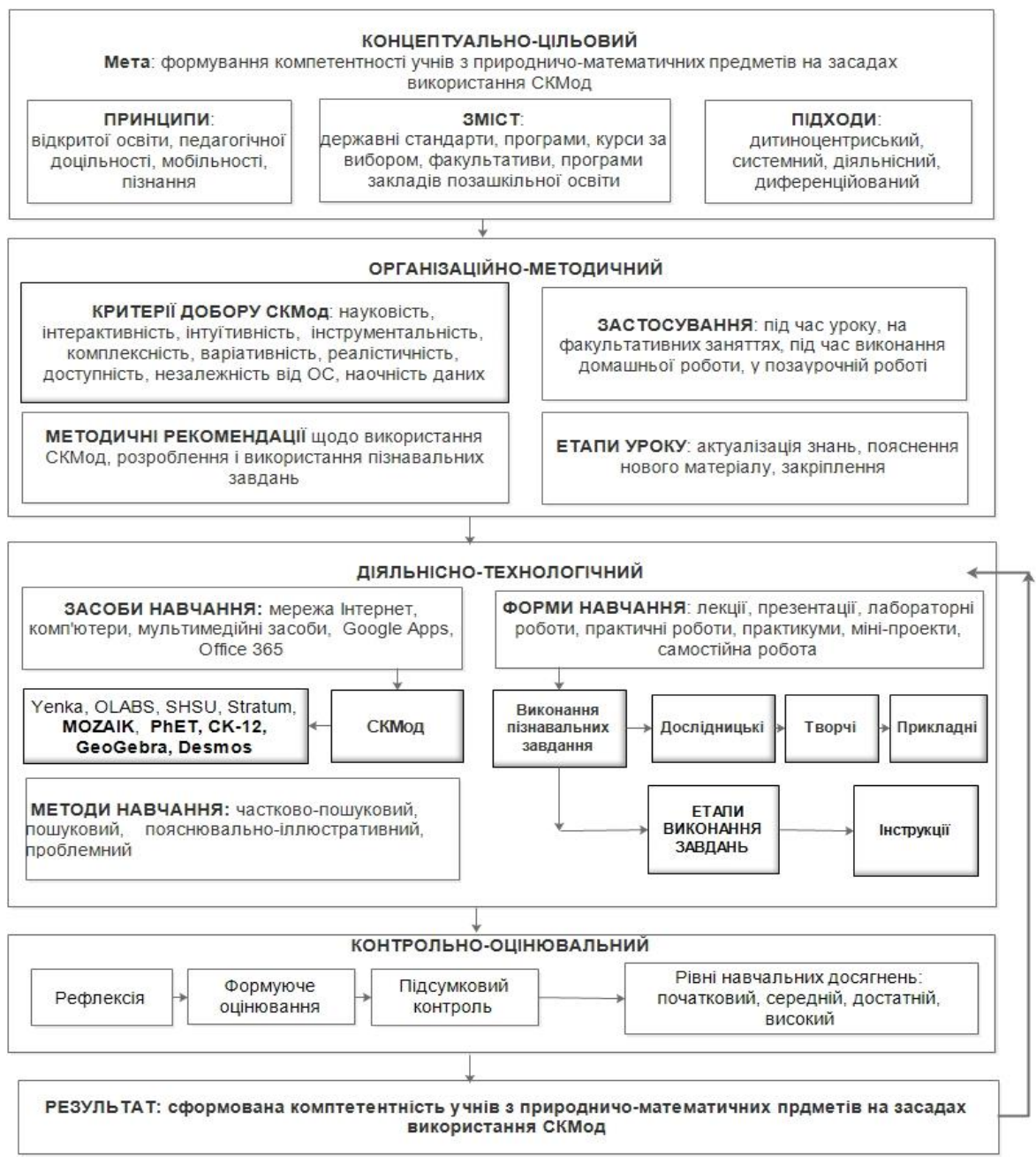

*Рис. 1. Модель використання системи комп'ютерного моделювання для формування компетентностей учнів*

Головна мета нової української школи – сформувати компетентного випускника, який буде здатним застосувати отримані знання у житті, насиченому цифровими засобами комунікації, управління, отримання освіти, ведення бізнесу. Відповідно до поставленої мети здійснюється реформування загальної середньої освіти з урахуванням потреб цифрового суспільства. Серед потреб цифрового суспільства суб'єкти освітнього процесу займають провідне місце – зростають вимоги до педагогів, зокрема до їх цифрової компетентності.

У Європейській структурі цифрової компетентності педагогів (DigCompEdu) деталізується науково обґрунтована структура, що роз'яснює значення «компетентний у цифровій сфері», зокрема компетентності учнів визначені як наскрізні і предметні. Розвиток цифрової компетентності учня спрямовано за такими напрямками: інформаційна грамотність, робота з даними, комунікація, використання електронного контенту, відповідальність, розв'язання проблем (https://ec.europa.eu/jrc/en/digcompedu).

Такий підхід до формування цифрової компетентності учнів дасть можливість широко впроваджувати в систему освіти технології пов'язані з використанням СКМод, що сприятиме ефективному формуванню предметних компетентностей учнів, зокрема з природничо-математичних предметів.

Використання СКМод на предметах природничо-математичного циклу має здійснюватися на принципах:

*відкритої освіти* – система комп'ютерного моделювання має бути відкритою для всіх учасників освітнього процесу (учнів, педагогів). На її використання не впливають симпатії або навпаки, антипатії, власний настрій чи стан;

*мобільності* – доступ до системи комп'ютерного моделювання має здійснюється будь-де і будь-коли;

*педагогічної доцільності* – використання системи комп'ютерного моделювання має відповідати цілям уроку або освітнього процесу;

*пізнання*, що є методологією в обґрунтуванні набуття учнями знань під час використання системи комп'ютерного моделювання. У центрі досліджуваних систем комп'ютерного моделювання знаходиться учень як член соціуму, суб'єкт, мовна особа.

У процесі використання СКМод учителі мають застосовувати низку підходів, зокрема:

*дитиноцентриський* – спрямовувати освітній процес на розвиток особистості учня, закладених природою обдарувань, його пізнавальної діяльності, що є основоположним фактором у розвитку когнітивних здібностей особистості;

*системний* – визначати навчання як цілеспрямовану творчу діяльність учня, розглядати зв'язки між метою, завданнями, змістом, формами, методами навчання у взаємодії компонентів педагогічного процесу;

*діяльнісний* – спрямовувати організацію діяльності учня на використанням СКМод, використання якої активізувало б його пізнавальну діяльність, спілкування та процеси саморозвитку;

*диференційований* – полягає у забезпеченні прав обдарованих дітей та дітей з різними функціональними обмеженнями на отримання доступу до якісних освітніх послуг, зокрема до здійснення індивідуальної освітньої діяльності.

Зміст навчального матеріалу має відповідати державним стандартам, програмам, курсам за вибором, факультативам, програмам закладів позашкільної освіти. Зазначимо, що питання інтеграції цифрового контенту, зокрема СКМод, в освітній процес до кінця не визначено нормативними документами. Ще й досі відсутні організаційні та методичні освітні складові, як от: використання цифрових лабораторій, електронних освітніх ресурсів, електронних підручників, комп'ютерних моделей, що потребує модернізації навчальних (освітніх) програм.

Нині існує низка систем комп'ютерного моделювання, що пропонується виробниками для застосування в освітньому процесі, більшою мірою зарубіжні (Yenka (www.yenka.com), Olabs (http://www.olabs.edu.in), Stratum (http://www.stratum.ac.ru), Mozaik (www.mozaweb.com), Phet (tps://phet.colorado.edu), СК-12 (https://www.ck12.org), GeoGebra (https://www.geogebra.org), Desmos (https://www.desmos.com/)). Учителям бажано здійснювати критичний відбір таких ресурсів. Пропонуємо критерії добору систем комп'ютерного моделювання, використання яких спрямовується на формування природничо-математичних компетентностей учнів:

*Варіативність* – передбачено можливість формулювати і вирішувати завдання декількох варіантів та рівнів складності.

*Доступність* — доступ до комп'ютерної моделі може здійснюватися будь-де і будь-коли.

*Інструментальність* – мають цифровий інструментарій для проведення низки вимірювань.

*Інтерактивність* – зміни певних параметрів призводять до змін в системі комп'ютерного моделювання.

*Інтуїтивність* – керування не потребує докладних покрокових інструкцій користування або передбачає мінімальні інструкції і супровід.

*Комплексність* – вирішення/дослідження не однієї проблеми/задачі, а комплекс різних завдань.

*Наочність даних* – вбудовані анімаційні фрагменти, графіки і діаграми, що ілюструють перебіг явищ і процесів.

*Науковість моделей* – відображає реальний світ, відповідає природничим законам, працює на «граничних» випадках.

*Незалежніст*ь від операційної системи і пристрою - перегляд і управління на різних цифрових пристроях (ПК, мобільних телефонах, планшетах) з різними операційними системами.

*Реалістичність* – наближене до реального життя учнів, їх оточення.

Впровадження СКМод в освітній процес має підкріплюватися методичними матеріалами, які допоможуть вчителю отримати відповідь на загальні методичні й організаційні питання.

До організаційних заходів використання системи комп'ютерного моделювання потрібно віднести визначення часу її використання: під час уроку, на факультативних заняттях, під час виконання домашньої роботи або використати в позаурочній діяльності (на гуртках, під час підготовки робіт на конкурс Малої академії наук тощо).

Ефективність використання цифрового контенту залежить від майстерності вчителя інтегрувати його в конкретний етап уроку: під час актуалізації знань, пояснення нового матеріалу або закріплення.

Також треба звернути увагу на добір форм навчання учнів з використанням СКМод. Це можуть бути традиційні форми: лекції, презентації, лабораторні і практичні роботи, практикуми, міні-проекти, самостійна робота учнів. Але можна організувати і новітні форми, наприклад *дослідницька станція* (фізика, біологія) – працюють групи учнів, результати і висновки записують у спільний звіт (документ) потім йде обговорення, порівняння і узагальнення або *слідство ведуть знавці* (хімія) – групи учнів роблять припущення і шукають шляхи вирішення проблеми, проводять експеримент потім йде презентація результатів і обговорення.

Оскільки основний фокус ми робимо на розвиток пізнавальних здібностей учнів, тому проектування пізнавальних завдань, визначення етапи їх виконання, розробка інструкцій, стають важливою складовою впровадження СКМод в освітній процес (Литвинова, 2018).

Перевагу пізнавальним завданням ми будемо віддавати таким, які можна реалізувати на рівні навчального заняття: уроку, домашнього завдання, практичної роботи або навчального проекту. Визначимо основні рівні пізнавальних завдань за характером мислення учня:

*дослідницьке* – організація та здійснення дослідження проблеми на засадах навчання;

*творче* – узагальнення даних та розробка власних моделей, розв'язків;

*прикладне* – застосування актуальних знань, умінь і компетентностей для вирішення навчальної або життєвої проблеми.

Важливого значення набуває проектуванні пізнавальних завдань і технологія їх застосування в освітньому процесі. Цей аспект детально висвітлено в праці (Литвинова, 2018)

Зупинимося ще на такому етапі уроку як підведення підсумків. Нині у цьому процесі активну участь можуть взяти учні. Ми пропонуємо вчителям застосовувати рефлексію. Рефлексія – це самоаналіз (Бусел, 2005).

Рефлексію класифікують її за функціональним змістом:

*рефлексія настрою* та емоційного стану – використовується на початку уроку і в кінці діяльності учнів для перевірки їх емоційного стану;

*рефлексія діяльності* – використовується для осмислення способів та прийоми роботи з навчальним матеріалом, пошуку раціональних варіантів;

*рефлексія змісту навчального матеріалу* – використовується для виявлення рівня усвідомлення змісту, вивченого на уроці (Орбан-Лембрик, 2005).

Доречними для підведення підсумків будуть картки, наприклад із зображенням емоційних облич (смайлики). Або низка прийомів:

«Ромашка Блума» – шість запитань розроблених на засадах таксономія Блума (знання, розуміння, застосування, аналіз синтез, оцінювання);

«Рибацька сітка» – учні виписують властивості вивченого об'єкта в таблицю;

«Дерево припущень» – коріння/гіпотеза, гілки/шляхи розв'язання, стовбур/результат і т.д.

Рефлексивна контрольно-оціночна діяльність при організації навчального процесу важлива, оскільки не тільки аналізує результати роботи учнів, але й сам процес роботи, що у свою чергу, передбачає включення кожного учня у взаємоконтроль і взаємооцінювання (Тихоненко, 2016).

Підвищити ефективність контролю та оцінювання знань учнів можна за допомоги формуючого оцінювання.

Формуюче оцінювання застосовується вчителями для отримання даних щодо поточного стану засвоєння знань учнями конкретної теми і визначення найближчих кроків щодо їх покращення.

Особливості формуючого оцінювання полягають в тому, що оцінюється робота учня у досягнення цілей навчання, а не його особистість; пропонується чіткий алгоритм визначення оцінки, зрозумілий учню; увага акцентується на персональному прогресі учня, а не на оцінці.

Форми проведення формуючого оцінювання учнів пропонуються такі: рефлексивні техніки (сигнали рукою, картками) для з'ясування і виявленням складних питань; уточнюючі питання; аналітичні питання; міні-тести; перевірка творчих робіт з метою виявлення помилок тощо.

Підсумкове оцінювання застосовується для визначення якості засвоєння учнями навчального матеріалу за конкретною темою.

Погоджуємося з думкою науковців які зазначають, що інструментами фіксаціїї оцінювання навчальних досягнень учнів може бути особистий профіль компетенцій, особисте віртуальне портфоліо, стрес-тест віртуального світу або цифрової моделі.

З метою заохочення та мотивування учнів до навчально-пізнавальної діяльності застосовуються змагальні ігрові моделі (гейміфікація), системи управління репутаційним капіталом, превентивне управління результатом (системи прогнозування досягнень), ігрові адаптивні моделі, системи моніторингу стану (що відстежують емоційний стан учнів). (Пінчук&Соколюк, 2018).

Основними функціями контролю навчальних досягнень учнів є: мотиваційна, діагностувальна, коригувальна, прогностична, перевірочна, розвивальна, виховна.

Оцінювання навчальних досягнень учнів має здійснюватися відповідно до Наказу МОН України від 13.04.2011 № 329 «Про затвердження Критеріїв оцінювання навчальних досягнень учнів (вихованців) у системі загальної середньої освіти» за яким критерії визначають загальні підходи до визначення рівня навчальних досягнень учнів відповідно до знань, умінь і навичок учнів та визначаються показником оцінки в балах. Основними критеріями оцінювання навчальних досягнень учнів є: *систематичність (*засвоєння навчального матеріалу в його логічній послідовності); *повнота* (обсяг знань про кількісні та якісні характеристики об'єкту); *міцність* (безпомилковість відтворення і збереження в пам'яті учня вивченого матеріалу); *оперативність (*уміння учня застосовувати знання про об'єкт у типових умовах); *гнучкість* (уміння учня використовувати знання про об'єкт у нетипових умовах).

Критерії реалізуються в нормах чотирьох рівнів досягнень: початковий, середній, достатній, високий. При визначенні рівня навчальних досягнень учнів враховуються: характеристики відповіді (правильність, логічність, обґрунтованість, цілісність); якість знань; сформованість загальнонавчальних та предметних умінь і навичок; рівень володіння розумовими операціями: вміння аналізувати, синтезувати, порівнювати, класифікувати, узагальнювати, робити висновки тощо; вміння виявляти проблеми та розв'язувати їх, формулювати гіпотези; самостійність оцінних суджень.

Застосування новітньої СКМод сприятиме як активізації навчальної діяльності учнів, так і формуванню їх компетентностей. Відповідно до Закону про освіту *компетентність* - динамічна комбінація знань, умінь, навичок, способів мислення, поглядів, цінностей, інших особистих якостей, що визначає здатність суб'єкта успішно соціалізуватися і здійснювати навчальну діяльність. Тому усі зазначені заходи сприятимуть формуванню і розвитку компетентності учнів з природничо-математичних предметів.

**4. ОБГОВОРЕННЯ**

Обізнаність вчителів щодо наявних СКМод дуже низька (Литвинова, 2018). Тому, виникає потреба в інформуванні і навчанні вчителів-предметників використанню СКМод в освітньому процесі: під час уроку, в організації роботи учнів із засобами інформаційно-комунікаційних технологій в позаурочний час (на факультативах і гуртка), у здійсненні контролю виконання пізнавальних завдань та самостійної роботи.

Впровадження моделі використання СКМод в систему загальної середньої освіти потребує удосконалення змісту і включення до програми підвищення кваліфікації та підготовки майбутніх вчителів таких тем, як: поняття про комп'ютерне моделювання, електронні освітні ресурси, структура пізнавального завдання, застосування моделі використання СКМод для навчання учнів природничо-математичним предметам, робота з мережевими програмами для організації опитування та контролю знань учнів.

Основними напрямами формування компетентності майбутнього вчителя природничо-математичних предметів з використання системи комп'ютерного моделювання, як складової професійної підготовки є: формування освітнього середовища; організаційні аспекти використання СКМод; використання мультимедійного комплексу для роботи з класом; теоретичні засади розвитку систем комп'ютерного моделювання; теоретико-методичні засади проектування і використання пізнавальних завдань для різних вікових категорій; форми реалізації СКМод в освітньому процесі; методи навчання учнів з використанням СКМод; організаційно-методичні аспекти розробки уроку з використанням СКМод, електронних освітніх ресурсів; віртуальний кабінет вчителя (хмаро-орієнтовані технології); моніторинг рівня навчальних досягнень учнів; оцінювання уроку з використанням СКМод; оцінювання якості та добір комп'ютерних моделей.

Ці напрямки підготовки мають бути включені до навчальних планів закладів вищої освіти, що здійснюють підготовку майбутніх вчителів природничо-математичних предметів.

**5. ВИСНОВКИ ТА ПЕРСПЕКТИВИ ПОДАЛЬШОГО ДОСЛІДЖЕННЯ**

Системи комп'ютерного моделювання стають альтернативою звичайним малюнкам у паперових підручниках і плакатам. Імітаційні процеси, інтерактивні вправи різних типів, анімаційний супровід, формуюче оцінювання дають можливість кожному учню формувати компетентності і пізнавати світ з захопленням.

Використання СКМод урізноманітнює процес здобуття освіти, дозволяє перейти від пасивних до активних методів навчання, активізує навчально-пізнавальну діяльність учнів, дозволяє створити індивідуальну траєкторію розвитку для кожного учня у процесі вивчення природничо-математичних предметів.

Аналіз наукових праць дав можливість зробити висновок, що комп'ютерне моделювання сприяє розвитку інтелектуальних умінь, глибокому розумінню процесів, що моделюються; формуванню дослідницьких умінь, поглибленню знань і вмінь з інформатичних, фізичних, хімічних, біологічних та математичних дисциплін, удосконаленню навичок роботи з СКМод, отриманню нових знань на засадах діяльнісного підходу.

Вчитель-предметник, як координатор цього розвитку, має володіти новітніми технологіями навчання на засадах використання інформаційно-комунікаційних технологій. Удосконалення системи підготовки майбутніх вчителів-предметників потребує не тільки базового знання про комп'ютерну техніку, а й нові підходи формування освітнього середовища, використання новітніх технологій навчання, зокрема СКМод.

З викладеного вище можна зробити висновки, що особливостями моделі використання СКМод є: формування компетентностей учнів з природничо-математичних предметів як цілісної відкритої, динамічної системи; здійснення інноваційного освітнього процесу в умовах динамічних технологічних змін; забезпечення координації індивідуального розвитку кожного учня для успішної реалізації компетентнісного підходу; формування уявлення учнів про цілісність картини світу. СКМод стає засобом дослідження процесів і явищ живої та неживої природи. Комп'ютерне моделювання таких процесів сприятиме розробленню системи пізнавальних завдань для учнів на засадах науковості, доступності та наочності. Застосування системи пізнавальних завдань формуватиме в учнів предметні компетентності, уявлення про оточуючий світ і взаємозв'язки в природі.

Подальшого обґрунтування потребує розроблення методичних рекомендацій щодо застосування СКМод в освітньому процесі, що слугуватимуть науково-методичним забезпеченням навчання учнів предметів природничо-математичного циклу.

**Список використаних джерел**

**THE MODEL OF THE USE OF COMPUTER MODELING SYSTEM FOR FORMATION COMPETENCES OF NATURAL AND MATHEMATICAL SUBJECT STUDENTS**
**Lytvynova Svitlana**
*Institute of Information Technologies and Learning Tools of National Academy of Education Sciences of Ukraine*

**Abstract.**

***Problem formation***. *Lack of methodical support, low level of teachers' awareness of existing effective teaching technologies such as computer modeling does not allow students to form their own individual trajectory for development as well as their competence in natural and mathematical education. This is annually confirmed by the analytical reports of the Ukrainian center for education quality assessment, in particular, the results of external independent assessment of students studying natural –mathematical sciences which are published.*

***Materials and methods***. *Methods of pedagogical and methodical literature analysis as well as dissertation project analysis were used in the scope of this research; the generalization of domestic and foreign experience results was carried out; theoretical modeling of the use of the computer modeling system for the development of students' competences; as well as system analysis to determine the structural elements of the model of computer modeling system.*

***Research results***. *The model of use of the computer modeling system for developing students' competences in natural and mathematical subjects has been developed. The peculiarities of this model include its use on the basis of child-centric, systemic, activity, differentiated approaches with the purpose of forming natural and mathematical subject students' competences. An important result of the study is the justification of: CMODS selection criteria (variability, accessibility, tooling, interactivity, intuition, complexity, data visuality, scientific ground of models, independence, realism); forms and methods of applying CMODS in the educational process, allocating the aspect of cognitive tasks application (research, creative, applied) for the students competences formation; the justification of such forms of evaluation as reflection, formative evaluation, final*

*evaluation.*

***Conclusions****. Computer modeling systems become an alternative to conventional drawings in paper course books and posters. Imitation processes, interactive exercises of various types, animation support as well as formative assessment give each student the opportunity of developing competence and exploring the world with enthusiasm. The use of CMODS diversifies the process of obtaining knowledge, allows transfer from passive to active teaching methods, activates educational and cognitive activity of students, as well as the creation of an individual trajectory for development for each student in the process of studying natural and mathematical subjects.*

*The development of methodical recommendations on the use of CMODS in the educational process requires further justification; the methodical recommendations will serve as a scientific and methodical support for the formation of competences of students studying natural and mathematical subjects.*

***Key words:*** *computer simulation system; CMODS; modeling; secondary schools; natural and mathematical education; competence; cognitive tasks.*